\documentstyle[aps,multicol,epsf,epsfig,aps,rotate]{revtex}

\begin{document}

\draft

\title{The Approximate Invariance of the Average Number of Connections
 \\ for the Continuum Percolation of Squares at Criticality}

\author{Sameet Sreenivasan,$^\ast$ Don R. Baker,$^{\ast\dagger}$ Gerald
  Paul,$^\ast$ and H. Eugene Stanley$^\ast$}

\address{$^\ast$Center for Polymer Studies and Dept. of Physics,
Boston University, Boston, MA 02215 USA\\ 
$\dagger$Department of Earth and Planetary Sciences,
McGill University\\ 3450 rue University, Montr\'eal, QC H3A 2A7 Canada}

\date{24 May 2002}

\maketitle

\begin{abstract}

We perform Monte Carlo simulations to determine the average excluded area
$<A_{ex}>$ of randomly oriented squares, randomly oriented widthless
sticks and aligned squares in two dimensions. We find significant
differences between our results for randomly oriented squares and previous
analytical results for the same. The sources of these differences are
explained. Using our results for $<A_{ex}>$ and Monte Carlo simulation
results for the percolation threshold, we estimate the mean
number of connections per object $B_c$ at the percolation threshold for
squares in 2-D. We study systems of squares that are allowed random
orientations within a specified angular interval. Our simulations show
that the variation in $B_c$ is within 1.6\% when the
angular interval is varied from 0 to $\pi/2$.

\end{abstract}

\begin{multicols}{2}

\section{Introduction}
Continuum percolation has been of significant interest in the study of
porous media \cite{Bunde}. It offers important advantages over lattice
percolation due to the fact that the majority of systems encountered in
nature are not confined to a lattice and are therefore modeled more
appropriately using continuum systems \cite{Bunde,Ben-Avraham00,Sahimi94,SA94}.

When studying the transport properties in porous media, the connectivity 
properties of the spanning cluster at percolation threshold are
important. One measure of the connectivity is the mean number of
connections per site.
In the case of lattice percolation, Scher and Zallen
\cite{ScherZallen70} demonstrated the approximate dimensional invariance
of this quantity. The behavior of
the analogous quantity $B_c$ in continuum percolation systems has been
previously studied \cite{Alonetal90,Balbergetal84,Balberg87}.  
In the case of continuum percolation, the product of the critical
concentration $N_c$ of objects at the percolation threshold and the average
excluded area $<A_{ex}>$ gives the critical average
number of intersections per object
$B_c$\cite{Alonetal90,Balbergetal84,Balberg87}; 
\begin{equation}
B_c = N_c <A_{ex}>.
\label{equation1}
\end{equation}
The excluded area of an object is defined as the area around an object
into which the center of another similar object is not allowed to enter if
intersection of the two objects is to be avoided \cite{Onsager49}. In the case
of objects that are allowed random orientations in a specified angular
interval, one defines an average excluded area $<A_{ex}>$ , that is the
excluded area averaged over all possible orientational
configurations of the two objects. 
 It has been claimed \cite{Balbergetal84} that $B_c$
for percolating systems of differently shaped
objects lies within a bounded range : in 2-D, $3.57 \le B_c \le
4.48$. $B_c$ represents the connectivity in the spanning cluster and is
of interest as the invariance of $B_c$ would enable us to estimate the
percolation threshold $N_c$ using Eq.(\ref{equation1}), once $<A_{ex}>$
has been calculated.       

In the present work we focus on continuum percolation systems of squares in
2-D, in which the objects are allowed random orientations
within a specified angular interval. The motivation for the study of
such orientationally constrained systems comes from the geological
observation that fractures in rocks do not have random isotropic
orientations but are oriented within a more or less fixed angular
interval. For our system of squares there is one
angle $\theta$ that specifies the orientation of the object and we
constrain it to lie within $-\theta_\mu \le \theta \le \theta_\mu$. We
determine the percolation thresholds for different values
of the constraint angle $\theta_\mu$. Simulations are also performed to find
the excluded area for each case. Our results show that for a given
object shape $B_c$ is constant to within 1.6\% for squares 
independent of the value of the constraint angle $\theta_\mu$.

\section{Simulation Method for finding Excluded Area}

Here we describe the method used to determine the excluded area
 for a pair of objects that are allowed random orientations within
the angular interval $-\theta_\mu$ to $\theta_\mu$.
For rectangles  (squares being a
particular case)
$\theta_\mu = 0$ corresponds to the case where the objects are aligned
parallel to each other and $\theta_\mu = \pi/2$ corresponds to the random
isotropic case. We describe the algorithm for finding the excluded area
 of squares in 2-D. 

A square of unit side is placed with its center coinciding with the center
of a lattice of edge length $L = 5$. The lattice size is chosen to be
larger than the excluded area, but small enough to sufficiently minimise the
number of wasted trials and yield good statistics. The square is given an
orientation $\theta_i$, randomly chosen in the interval  $-\theta_\mu \le
\theta_i \le \theta_\mu$, with the reference axis. A second square is then
introduced into the lattice with its center randomly positioned in the
lattice. This square is given an orientation $\theta_j$ chosen randomly
from the same interval as for the first object. We then determine if the
two squares overlap. We repeat this procedure for $10^{9}$ trials and
record the number of times the two squares overlap. This number divided
by the number of trials is the probability that the two objects
overlap. The probability of overlap times the area, $L^{2}$, of the lattice
yields the average excluded area for a pair of squares oriented randomly
between  $-\theta_\mu$ to $\theta_\mu$. The method used to
determine the overlap of squares in 2-D is described in
detail in Ref.\cite{Bakeretal02}.

\section{Excluded Area Simulation Results}
 
  We determine the average excluded area for a unit square for
 different constraint angles. We also determine the excluded area of
 widthless sticks for the case of random isotropic orientations. In our
simulations the widthless stick is represented by a rectangle of
edge lengths $ 1 $ and $1 \times 10^{-12} $ . 
 Table 1 lists the Monte Carlo results for $<A_{ex}>$ obtained for the
 different objects studied for the case of random isotropic
 orientations. Table 2 shows the variation of $<A_{ex}>$ for squares
 with the constraint  angle. Our values for $<A_{ex}>$ for aligned
 squares are consistent with all previous results
 \cite{Balbergetal84}. Furthermore, our $<A_{ex}>$ values for
 randomly oriented widthless sticks are also consistent with earlier
 determinations \cite{Balbergetal84}; our slightly higher value of
 $<A_{ex}>$ compared to that of Ref.\cite{Balbergetal84} is
explained by the fact that our widthless sticks have a finite
width and thus are expected to exhibit a larger $<A_{ex}>$ than found
 analytically for the zero width limit.
However our $<A_{ex}>$ for the case of randomly aligned squares is
different from previous analytical results, our value
being $12$\% above that determined by Ref.
 \cite{Balbergetal84}. We propose a 
reason for this difference in the next section.

\section{Discrepancy with previous analytical determination of excluded area}

\begin{figure}
\centerline{
\epsfxsize=7.0cm
\epsfclipon
\epsfbox{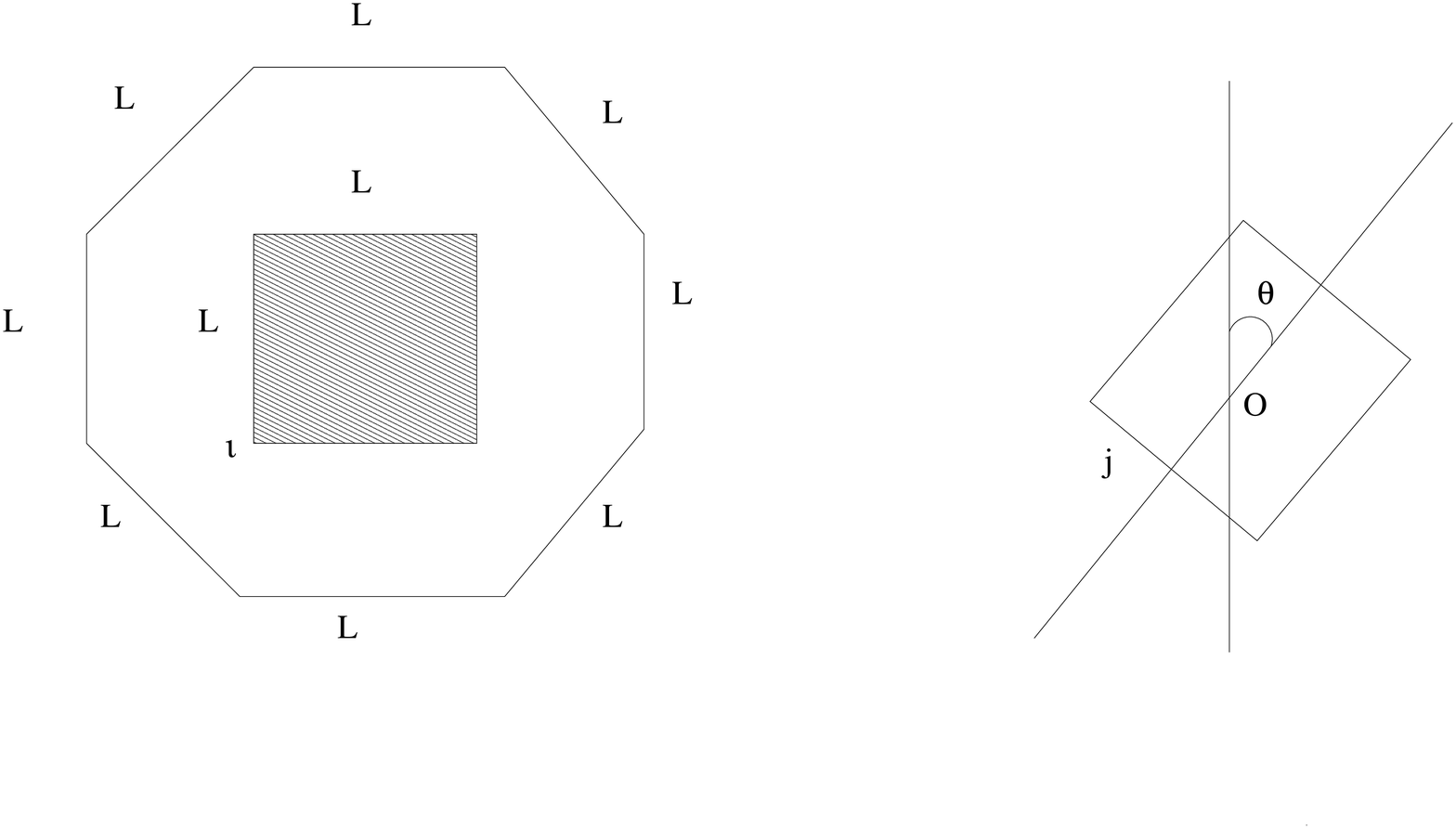}
}
\caption{ Procedure for determining the excluded area of two squares of
side L: the first square i (shaded), is kept fixed while the second
square j having orientation $\theta$ with respect to i, is moved around
i always keeping contact, and the locus of the center of j is found. The
area within the locus gives the excluded area for a given relative
orientation of the two squares.
}
\label{Figure1}
\end{figure}

We investigated the cause of the difference between our value of $<A_{ex}>$ for
squares in 2-D and the previous analytical result in
\cite{Balbergetal84} and found it to be the following:
In arriving at the expression for $<A_{ex}>$ Ref.\cite{Balbergetal84}
finds the excluded area for a pair of rectangles (Eq. 18) with
a given relative orientation $\theta$ (see Figure ~\ref{Figure1}). For
squares, using equation (Eq. 18) in Ref.\cite{Balbergetal84}, 
\begin{equation} 
A_{ex} = (\sin\theta + \cos\theta + 1)^2 - 2 \sin\theta \cos\theta,
\label{equation2}
\end{equation}
where
\begin{equation} 
\theta \equiv |\theta_i - \theta_j|,
\label{equation3}
\end{equation}
$\theta_i$ and $\theta_j$ being the individual orientations of the two squares.
Ref.\cite{Balbergetal84} then obtains the average excluded area by averaging the right hand side of
Eq.(\ref{equation2}) over all possible orientations of both objects,
 $-\theta_\mu \le \theta_i \le \theta_\mu$ and
$-\theta_\mu \le \theta_j \le \theta_\mu$ , using a uniform probability
distribution 
\begin{equation}
P(\theta_i)= P(\theta_j)= 1/2\theta_\mu.
\label{equation4}
\end{equation}
However, it appears Ref. \cite{Balbergetal84} overlooked the fact that Eq.(\ref{equation2}) holds only
for $0 \le \theta \le \pi/2$ (hence $0 \le \theta _\mu \le \pi/4$), since for
$\pi/2 \le \theta \le \pi$, the expression gives a value of $A_{ex}$ less
than the minimum possible value of 4 \cite{Balbergetal84}. Thus the
procedure of Ref.\cite{Balbergetal84} does not work
when the constraint angle $\theta_\mu$ is greater than $\pi/4$. The correct
result can be obtained by replacing $\theta$ in Eq.(\ref{equation2}) by
\begin{equation}
\theta' = \theta \bmod  (\pi/2), 
\label{equation5}
\end{equation}
  so that Eq.(\ref{equation2}) holds for all values of  $\theta_\mu$. For the
 random isotropic case $\theta_\mu = \pi/2$, using the modified
Eq.(\ref{equation2}) and integrating numerically we obtain $<A_{ex}> = 4.54647$ which is in close
 agreement with our Monte-Carlo simulation result.

We also calculate the values of $<A_{ex}>$ for squares with other values of
$\theta_\mu$ between $0$ and $\pi/2$ (Table 2). We
notice that the values of $<A_{ex}>$ are the same for $\theta_\mu = \pi/4$ and
$\theta_\mu = \pi/2$, which is true since the rotation of a square in
a particular configuration through an additional angle of $\pi/4$ yields the
same configuration. 

\begin{figure}
\centerline{
\epsfxsize=7.0cm
\epsfclipon
\rotate[r]{\epsfbox{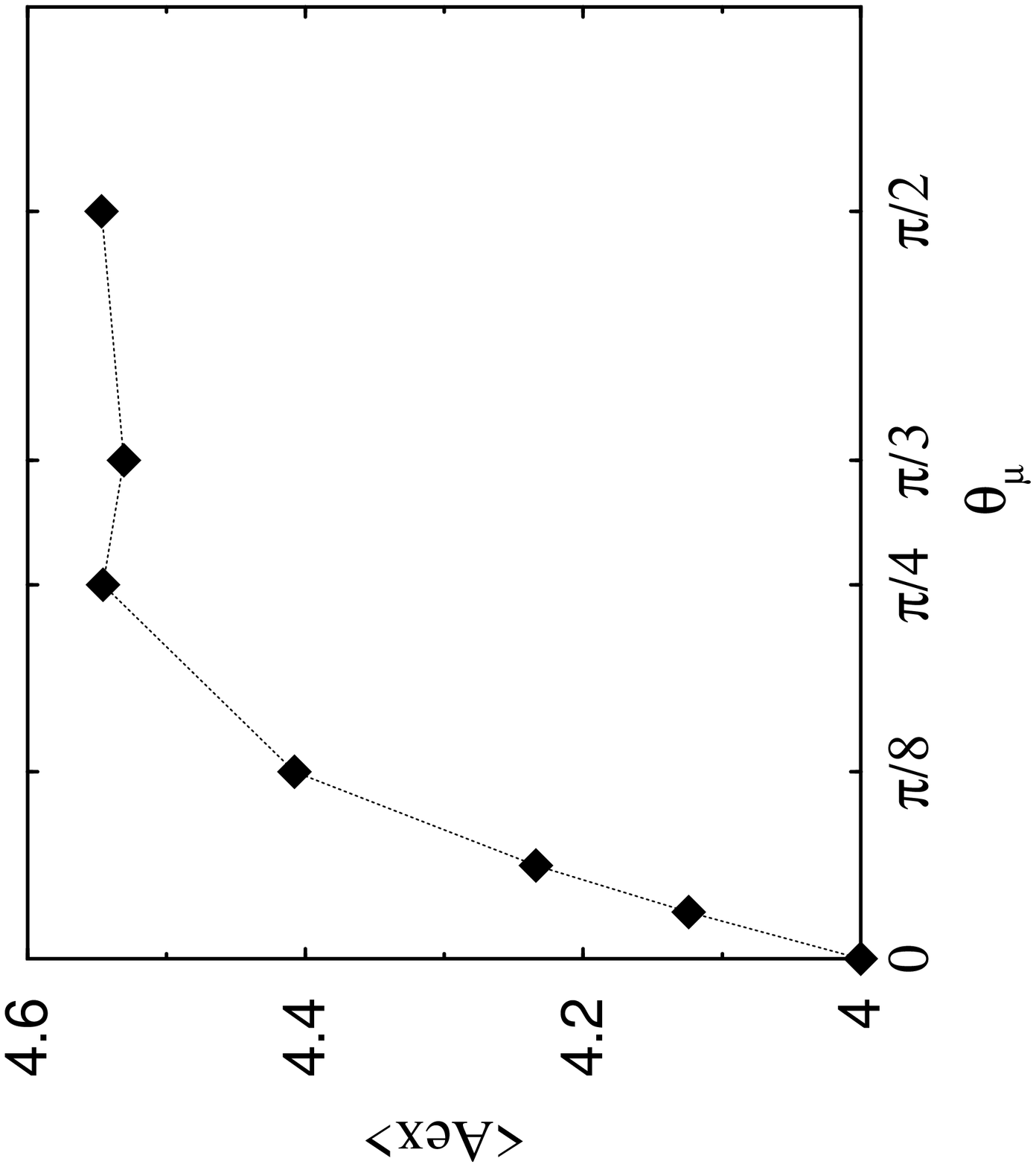}}
}
\caption{ Plot of the excluded area $<A_{ex}>$ for squares in 2-D versus
the constraint angle $\theta_\mu$. The uncertainties are smaller than the symbols.}
\label{Figure2}
\end{figure}

Note that the value of
$<A_{ex}>$ appears to decrease for $\theta_\mu > \pi/4$ , reaches a
local minimum near
$\theta_\mu=\pi/3$ and then
increases again till it reaches a maximum at $\theta_\mu = \pi/2$ (see Fig.~\ref{Figure2}). This can
be explained as follows. The case $\theta_\mu = \pi/4$ is
equivalent to the case of random isotropic orientation. Here the angle
$\theta = |\theta_i - \theta_j|$ can range from 0 to $\pi/2$. When
$\theta_\mu$ is greater than $\pi/4$, $\theta$ can take values greater than
$\pi/2$ which means that in addition to the configurations obtained for
$\theta_\mu=\pi/4$, there are other configurations for which the relative
orientation $\pi/2 \le \theta \le 2 \theta_\mu$. However the latter
configurations are, in fact, the same as those for  $\theta = (2\theta_\mu -
\pi/2) < \pi/2$ due to the symmetry of squares. Thus, the decrease in the
$<A_{ex}>$ between $\theta_\mu =\pi/4$  and $\theta_\mu = \pi/2$ can be
attributed to the increased probability of achieving 
configurations with smaller excluded areas.

\section{Percolation Thresholds}

\begin{figure}
\centerline{
\epsfxsize=7.0cm
\epsfclipon
\rotate[r]{\epsfbox{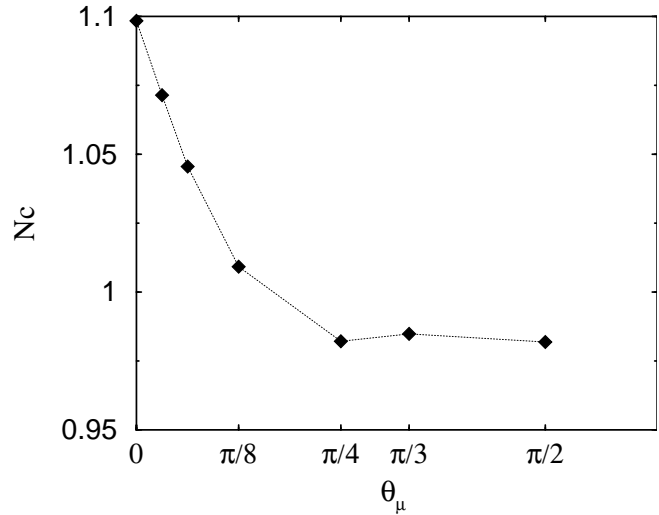}}
}
\caption{ Plot of the percolation threshold $N_c$ for squares in 2-D versus
  the constraint angle $\theta_\mu$. The uncertainties are smaller than the
symbols.}
\label{Figure3}
\end{figure}

Using the procedure of Ref \cite{Bakeretal02}, we perform Monte Carlo simulations for the determination of the percolation
threshold based upon the Leath method\cite{Leath76} and the methods Lorenz
and Ziff \cite{Lorenz01} used in their study of continuum percolation of
spheres. The only difference in our present simulations is
that the random numbers generated to fix the orientation of an object lie
within a specified angular range from $-\theta_\mu$ and $\theta_\mu$. We
determine the percolation thresholds of squares in 2-D for
different values of the constraint angle $\theta_\mu$. These results are
shown in Table 2. We also show the values of critical area fraction
$\phi_c$ . Fig.~\ref{Figure3} shows a plot of the percolation
threshold $N_c$ for squares in 2-D versus the constraint angle $\theta_\mu$. An interesting feature of the 2-D plot is that as we
increase the orientational freedom beginning from $\theta_\mu = 0$, $N_c$
drops until it reaches the value for $\theta_\mu = \pi/4$, then begins
increasing until it reaches a maximum and then falls again. This
behavior is expected if $B_c$ is to remain approximately invariant, as
we shall explain below.   

\section{Approximate Invariance of $B_c$}

Using the percolation thresholds and  excluded area obtained from
our Monte Carlo simulations, we find the average number\cite{Balbergetal84} of connections per
object at threshold $B_c$.
Table 2 shows the values of $B_c$ for squares for various values of the constraint angle
$\theta_\mu$. The change in $B_c$ in going from a constraint angle of
$\theta_\mu = 0$ to $\theta_\mu = \pi/2$ is less than $1.6$\%. Using the
old values of $<A_{ex}>$ \cite{Balbergetal84}, the
variation in $B_c$ is seen to be $\approx$ 9.5\%. We see
that the slight decrease in $<A_{ex}>$ between $\theta_\mu =\pi/4$  and
$\theta_\mu = \pi/2$ (see Fig.~\ref{Figure2}) is compensated by an
increase in the corresponding $N_c$ (see Fig.~\ref{Figure3}) to give an
approximately invariant $B_c$. The closeness of $B_c$ values for a given
system is striking and is consistent with the hypothesis that $B_c$ is  
approximately invariant for continuum percolating systems of a particular
shape.  

\section{Summary}

Our results show that the value of $B_c$ is approximately independent of
orientational constraints. The $B_c$ value
of a shape is indicative of the efficiency of the object in forming a
percolating cluster. Not only the magnitude of the excluded
area plays a part in the formation of connections, but also
the distribution of the average excluded area in space. This is easily seen
from the fact that both unit area discs and aligned unit area squares 
have $<A_{ex}>$ = 4 \cite{Balbergetal84}, but the percolation threshold of
aligned squares $\phi_c = 0.6666$ \cite{Bakeretal02} is lower
than that of discs $\phi_c=0.676339$ \cite{Quintetal00}. Our results suggest that the value
$B_c$ can be considered as a unique quantitative characteristic of a
shape and can therefore be useful in the prediction of the percolation
threshold as has been previously pointed out \cite{Alonetal90,Balbergetal84,Balberg87}.  

\subsubsection*{Acknowledgements}

 We thank R. Consiglio, A. Schweiger and P. Kumar for discussions and
 BP, NSERC and NSF for support.

\end{multicols}

\newpage

\begin{table}
\label{Table1}
\caption{Comparison of the average excluded area for
widthless sticks and squares in 2-D. The uncertainty in $<A_{ex}>$ is estimated as follows. The
reciprocal of the square root of the number of
Monte Carlo trials yielding intersection of the two objects is the
fractional uncertainty in the determination of $<A_{ex}>$. The product of the
fractional uncertainty and the estimated value of $<A_{ex}>$ is the uncertainty
in that value.}
\begin{tabular}{l|ccccc}
 \hline
Object & $<A_{ex}>$ for unit area object & Previous Result \\
\hline 
Widthless sticks & $0.6367\pm 0.0001$ & 0.6366\protect\cite{Balbergetal84} \\
Aligned squares & $3.9998\pm0.0003$ & 4\protect\cite{Balberg87} \\ 
Random Squares & $4.5466\pm 0.0004$  & 4.084\protect\cite{Balbergetal84}  \\ 
\end{tabular}

\end{table}

\newpage

\begin{table}
\label{Table2}
\caption{Critical area fraction, percolation threshold, average
excluded area and critical average number of connections per
object for squares in 2-D. Estimation of uncertainty in $\phi_c$ is described in \protect\cite{Bakeretal02}. }  
\begin{tabular}{l|ccccc}
 \hline
Constraint Angle & $\phi_c$ & $N_c$ & $<A_{ex}>$ & $B_c$ \\
\hline 
$\pi/2$ & $0.6254\pm 0.0002$ & 0.981896 & $4.5466\pm 0.0004$ & 4.464 \\
$\pi/3$ & $0.6265\pm 0.0005$ & 0.984837 & $4.5309\pm 0.0004$ & 4.462 \\
$\pi/4$ & $0.6255\pm 0.0001$ & 0.982163 & $4.5459\pm 0.0004$ & 4.465 \\
$\pi/8$ & $0.6355\pm 0.0005$ & 1.009229 & $4.4076\pm 0.0004$ & 4.448 \\
$\pi/16$ & $ 0.6485\pm 0.0005$ & 1.045546 & $4.234\pm 0.0004$ & 4.443 \\
$\pi/32$ & $0.6575\pm 0.0005$ & 1.071484 & $4.1240\pm 0.0004$ & 4.419 \\
0 & $0.6666\pm 0.0004$ & 1.098412 & $3.9998\pm0.0003$ & 4.394 \\ 
\end{tabular}
 
\end{table}

\newpage

\end{document}